\documentclass[acmsmall,screen,nonacm,review=false,timestamp=false]{acmart}

\newcommand{\Autoref}[1]{%
  \begingroup%
  \def\chapterautorefname{Chapter}%
  \def\sectionautorefname{Section}%
  \def\subsectionautorefname{Subsection}%
  \autoref{#1}%
  \endgroup%
}

\AtBeginDocument{%
  }

\setcopyright{acmcopyright}
\copyrightyear{2023}
\acmYear{2023}
\acmDOI{XXXXXXX.XXXXXXX}

\acmConference[RecSys '23]{the 17th ACM Conference on Recommender Systems}{September 18--22,
  2023}{Singapore}
\acmPrice{15.00}
\acmISBN{978-1-4503-XXXX-X/18/06}

\begin{document}

\title[RecSys Challenge 2023]{RecSys Challenge 2023: From data preparation to prediction, a simple, efficient, robust and scalable solution}

\author{Maxime Manderlier}
\email{maxime.manderlier@umons.ac.be}
\orcid{0000-0002-5641-9818}
\affiliation{%
  \institution{University of Mons (UMONS)}
  \streetaddress{Rue de Houdain 9}
  \city{Mons}
  \country{Belgium}
  \postcode{7000}
}

\author{Fabian Lecron}
\affiliation{%
  \institution{University of Mons (UMONS)}
  \streetaddress{Rue de Houdain 9}
  \city{Mons}
  \country{Belgium}
  \postcode{7000}
}
\email{fabian.lecron@umons.ac.be}
\orcid{0000-0002-6516-0086}

\renewcommand{\shortauthors}{Manderlier et al.}

\begin{abstract}
	The RecSys Challenge 2023\footnote{\url{http://www.recsyschallenge.com/2023/}}, presented by ShareChat, consists to predict if an user will install an application on his smartphone after having seen advertising impressions in ShareChat \& Moj apps. This paper presents the solution of 'Team UMONS' to this challenge, giving accurate results (our best score is 6.622686) with a relatively small model that can be easily implemented in different production configurations. Our solution scales well when increasing the dataset size and can be used with datasets containing missing values.
\end{abstract}

\ccsdesc[500]{Information systems~Decision support systems}


\keywords{online advertising, neural networks,  missing values, embeddings, binary classification}


\maketitle

\section{Introduction}

In the ShareChat \& Moj apps, ads are shown to users and invite them to install applications on their smartphone. By collecting data about the users and the ads we might be able to predict if one ad will conduct the user to install the advertised application on his smartphone. In this challenge, we were given a dataset of impressions (ads showed to users) and the labels 'is\_installed' and 'is\_clicked'. The challenge is evaluated based on prediction of the 'is\_installed' (Log-Loss is the evaluated metric) label but prediction of the second label is also important for further analysis. We will show how to predict these two important output variables.
\Autoref{data_preparation} describes how data is prepared, \autoref{model} describes our predicting model, \autoref{experiments} discusses about the achieved results, metrics, and observations, and \autoref{conclusion} concludes this work. The code for predicting the results of the challenge as well as the code that was used to compute metrics described in \autoref{experiments} are available in our github repository\footnote{\url{https://github.com/MaximeUM/RecSysChallenge2023}}.

\section{Data preparation}
\label{data_preparation}

As any machine learning practitioner should know, the best model will give bad results if it is fed bad data. Data preparation takes most of the time in many cases and includes some steps as handling missing values, normalization, feature selection, etc. We show in this section what steps we apply to our data.

As described by ShareChat RecSys Team, the dataset consists of records that capture user and ad features (categorical, binary, numerical) and whether a click and/or install was generated by the user. Description of the features is not given and therefore cannot be used to understand the data. With features descriptions, other modeling choices could be made and probably improve the quality of the predictions. In particular, we could model users and ads separately if we had information on which features correspond to the users and which correspond to the ads. We split the features by categories that we handle separately. Each one of them receive adequate data preparation steps as showed in \autoref{cat}, \autoref{bin} and \autoref{num}.

\subsection{Categorical data}
\label{cat}

There is 30 different categorical features ('f\_2' to 'f\_32'). First of all, we remove 'f\_7' from the data as that feature has one unique value (in the proposed dataset). It will therefore add no information and can be ignored. Some values are missing (NaN) and values exist in the test set but are never seen in the training set. These categorical values will be handled by embedding layers in our model (see \autoref{model}) that will allow us to use the value 0 for 'no information' and avoid training on that information. Thanks to that, estimating missing values for the categorical features is not necessary. In the columns, numbers are associated to the features but are not continuous. We will re-code the information to have each feature (column) containing numbers from 0 to n where n is the number of distinct values for that feature. That step is necessary as the embedding layers need continuous integer values. NaN are replaced by 0 and values in the test set that don't exist in the train set are also replaced by 0 (treated as missing as we never saw them).

\subsection{Binary data}
\label{bin}

9 different binary features ('f\_33' to 'f\_41') are given in the dataset. There are no missing values for these features (train set and test set). If we were confronted to a scenario with missing values, we would have replaced the missing values with coherent values as shown in \autoref{numerical_data}.
Being binary values, all values are between 0 and 1 for each feature and normalization is not necessary. No (further) treatment is needed and these data will be directly given as inputs to the model. 

\subsection{Numerical data}
\label{num}
\label{numerical_data}

Finally, the dataset contains 38 numerical features ('f\_42' to 'f\_79'). This category of data will be the one for which we really need to apply data preparation steps.

\subsubsection{Missing values estimation}
This part of the dataset contains missing values. One way to handle it would be to remove each row containing a NaN value (and remove the corresponding row from the binary data and categorical data). Doing that means that we would loose all the features for only one missing feature. Another idea would be to use masking in the neural network to avoid training for unknown values. However, proceeding like that doesn't allow the model to get a full representation of the inputs. Moreover, the problem of missing values is pushed to the next step (fitting a model) and will therefore not allow us to easily change from one model to another. Well done data preparation should allow us to use other models in \autoref{model} without much change.
As discussed in \cite{mv1,mv2,mv3,mv4}, better ways to handle missing values are available to us and will be used in this work.

In our case, we will replace the missing values by doing estimations based on the features of the dataset. Different methods are tested (replacement by the mean, by the median, by zero or by using the other features). The last method gave the more robust results and is the more elaborate. For each feature, we represent that feature as a function of the other features (only the numerical features here). Each time we encounter a missing value, we use the function defined for the feature to predict the value based on values of the other features. This method supposes that the variables are not independent and that similar values on some attributes should reflect a similar value on the attribute we try to estimate. The implementation we used is given by \textit{scikit-learn\footnote{\url{https://scikit-learn.org/stable/modules/generated/sklearn.impute.IterativeImputer.html}}}.

\subsubsection{Data scaling}
Also, for these features the ranges of values are really diverse. For instance, 'f\_74' ranges from 0 to 0.1157 while 'f\_64' ranges from 0 to 415,706,830,424. If the model is trained with non-normalized data, it would be difficult for features as 'f\_74' to get importance (these features would merely be ignored). All features will be normalized between 0 and 1 using the min-max normalization given by the following formula:
\begin{center}
$x_i = \frac{x_i - min(X)}{max(X) - min(X)}$
\end{center}
where $X$ represents one feature vector (e.g. 'f\_74') and $x_i$ one value of this vector. The formula is applied to each value of the feature vector and repeated for all feature vectors.

\section{Fitting the model}
\label{model}

\begin{figure}
\includegraphics[width=0.7\textwidth]{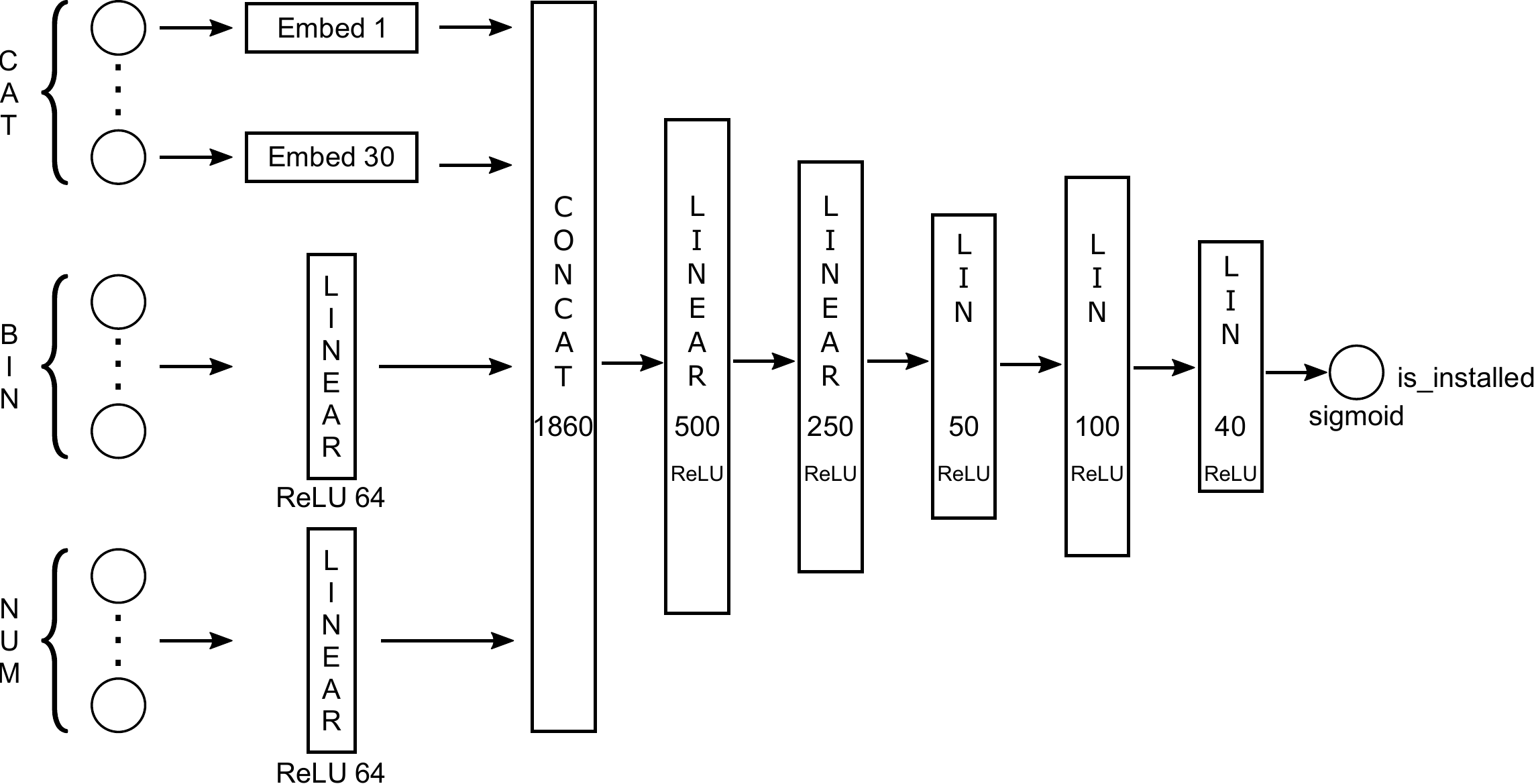}
\caption{Predict 'is\_installed'}
\label{one_output}
\end{figure}

\begin{figure}
\includegraphics[width=0.7\textwidth]{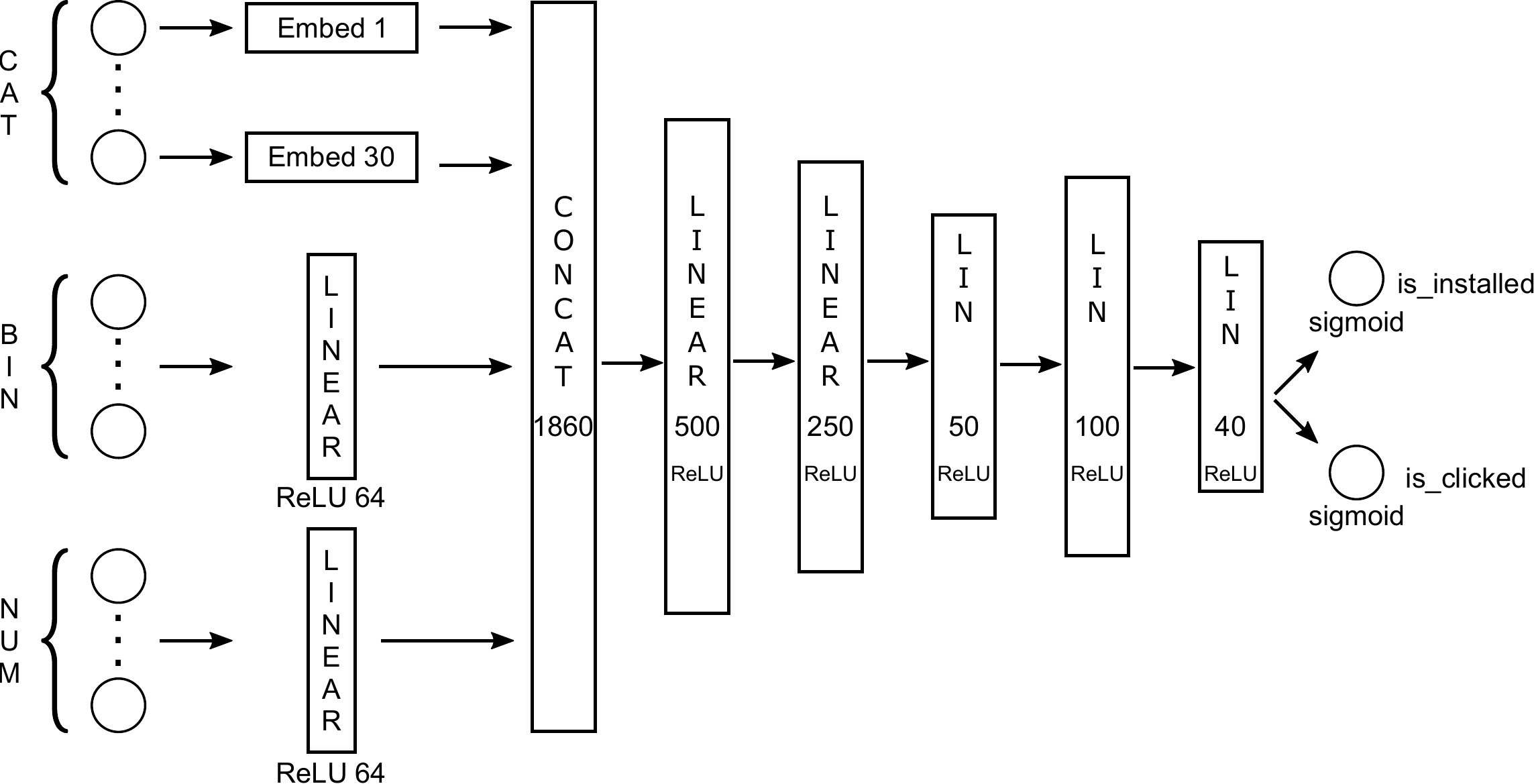}
\caption{Predict 'is\_clicked' and 'is\_installed'}
\label{two_outputs}
\end{figure}

For this challenge, as evaluation was associated to the Log-Loss, a model that is good at predicting probabilities was necessary. For that reason, we decided to use a deep neural network with a final \textit{sigmoid} function. This choice is also motivated by the use of neural networks in other works related to online advertising \cite{fire_exploring_2017, zhai_deepintent_2016, zhang_deep_2016, qu_product-based_2016}, proof that this kind of model is suitable for the given problem.
Using the neural network was also the opportunity for us to use each type of data (categorical, binary and numeric) in a different way and without doing too much transformation. In other words, the neural network is appropriate to use with multiple inputs (and also multiple outputs as we will see).
We describe the design of our models (\autoref{one_output} and \autoref{two_outputs}) in the following of this section.

\subsection{Defining the inputs}

\subsubsection{Categorical data}

The number of distinct values differs from one categorical feature to another one. One easy way to proceed would be to define, for each feature, n input neurons (with n being the number of distinct values for that feature). Some features have a lot of distinct values and each time we have a value of 1 for the corresponding neuron we have n-1 zeros input neurons. For that reason, we will use embedding layers. An embedding layer encodes the values of the features given m neurons (m is to be chosen). Instead of having one neuron for each distinct value, we use a combination of a smaller number of values to encode the same information (compression). If we have $n<=256$ distinct values (excluding the value 0 that is reserved for missing values), we choose $n$ as the number of outputs of the embedding layer as it would make no sense to choose more output nodes than there is distinct values. For features with more than 256 distinct values, we encode the information using 256 neurons.

\subsubsection{Binary data}

For the binary data, we use one input neuron per feature. After that input, we connect a 64 neurons linear layer with \textit{ReLU} activation function. 

\subsubsection{Numerical data}

For numerical data, the same logic as for binary data is followed, one input neuron for each numerical feature. We also connect these neurons to a layer of 64 neurons with \textit{ReLU} activation function.

\subsubsection{From the inputs to the output(s)}

Each categorical feature embedding, the binary linear layer and the numerical linear layer are concatenated together. Some linear hidden layers are piped (with \textit{ReLU} activation functions). Finally, one (\autoref{one_output}) or two (\autoref{two_outputs}) output layers are proposed at the end of our model. A first version with only one output layer is proposed to predict 'is\_installed'. The idea behind that is that it is the only output evaluated in the challenge and it accelerates the training for different configurations during the experiments. We then propose a version of the algorithm that contains two outputs ('is\_clicked' and 'is\_installed'). In \autoref{experiments}, we will investigate how this second output affects the predictions.

\section{Experiments}
\label{experiments}

In this section, we describe the experiments and the achieved results of our model. First, we discuss about the training and validation procedure and secondly about how we obtained the results for the challenge. We then compare the models defined in \autoref{model}.
All our experiments were made using \textit{TensorFlow} for the model (neural network) and \textit{scikit-learn} for data preparation and evaluation.

\subsection{Validation procedure}

Neural networks are prone to overfitting \cite{Ying_2019}. If we train the model for too much epochs, the loss will surely decrease. However, the model will perform very poorly on the test dataset as it will only be able to predict what it has already seen. 

We randomly split our training data into a training set and a validation set (ratio 3:1). The loss on the validation set will be used to monitor training and to keep the weights for which the validation loss is the best (lowest). We monitor the loss (binary crossentropy) that is the equivalent of the Log-Loss metric used for evaluation in this challenge. If this loss stops decreasing, training will be stopped (after some epochs to be sure that it was not a momentary increase) and the best model (lowest validation Log-Loss) kept for inference.

\subsection{Metrics}
\label{metrics}

In this challenge, solutions are evaluated based on the Log-Loss, meaning that the probability (confidence) of our predictions is important. The probabilities are for sure important for business because different actions could be pursued based on the confidence we have in our model. For example, if the probability of installing the application is 0.51, one more ad (or more) could be showed to the user to increase the chances that he installs the app. On the contrary, if the probability is high it might not be useful to show another ad. However, the model can be not so good to predict probabilities but still very good at predicting binary answers. We will monitor the 'no information rate' (NIR), the accuracy for a model always predicting the majority class; the accuracy; the true positive rate (TPR) or recall; the true negative rate (TNR) or specificity; the precision and the F\textsubscript{1} Score that are standards metrics used for evaluation of binary classification problems \cite{metrics,raschka2014overview}. 

The results obtained on the dataset are given in \autoref{metrics_is_installed}. We show results for the scenario where we split the dataset in a training and validation set, only way for us to compute metrics on data that were not seen by the model. The test set cannot be used as we don't have the associated labels and will only be used to submit predictions for this challenge. We also give the values of the metrics for the case in which we use all the data for training as it is the one we use for predicting labels of the test set. For training a model on the whole dataset, we use the same number of epochs that were found by the train/val experiment (3 epochs).

\begin{table}
\caption{Metrics for the training set and validation set (predict 'is\_installed')}
\label{metrics_is_installed}
\begin{tabular}{lccc}
\toprule
                  & Training set (75\%) & Validation set (25\%) & Training set (100\%)  \\
\midrule
Log-Loss          				&      0.3106         &  0.3177       &    0.3088        \\
NIR               				&      0.8258         &  0.8265       &    0.8260        \\
Accuracy          				&      0.8237         &  0.8207       &    0.8266       \\
TPR (Recall)      				&      0.5425         &  0.5324       &    0.5318        \\
TNR (Specificity) 				&      0.8830         &  0.8812       &    0.8888       \\
Precision         				&      0.4943         &  0.4849       &    0.5018       \\
F\textsubscript{1} Score       &      0.5173         &  0.5075       &   	0.5164        \\
\bottomrule
\end{tabular}
\end{table}

\subsection{Results interpretation}
\label{results_interpretation}
The NIR value in our case corresponds to always predict 0. We can observe that in the scenario where we split the data, the NIR is better than the accuracy (training set and validation set); the accuracy is slightly better than the NIR when we use the full dataset to train the model. It might then seem useless to use a model for prediction facing this reality. However, it might also seem really important to predict well the positive class, meaning to know when a combination of user and ad characteristics will lead to installing an application on the smartphone. If we always predict zeros, the true positive rate (TPR) is equal to 0. By using our model, we can observe that the TPR is higher than 0.5, meaning that we are able to detect more than half of the positive instances. The precision is also close to 0.5 meaning that half the time we say that a combination of features will lead to an install, it will. The TNR is much higher, meaning that it is easier to detect the instances of the negative class (majority class). Our model is then able to distinguish between positive and negative instances, which is good news. 
The Log-Loss is given in \autoref{metrics_is_installed} but is difficult to interpret as it is. We can compare it to other models developped for this challenge.

\subsection{Increasing Recall and Precision}

If increasing the recall is the goal we want to pursue, maybe that monitoring the val\_loss is not what we should do. We show in \autoref{metrics_more_epochs} that if we train the model for more than three epochs, we get worse results for accuracy and log-loss (which is not a good news from the challenge perspective) but increase other important metrics. As reflected in \autoref{metrics_more_epochs}, increasing TPR tends to decreases Precision. F\textsubscript{1} Score regroups these metrics and doesn't significantly decreases (but still is). From a business perspective, the most important metrics should be defined and used to monitor training. These important metrics are likely to evolve with time as new considerations will be taken into account. Our model is easily trainable for the metrics chosen in the future.

\begin{table}[]
\caption{Metrics for some training scenarios ('is\_installed')}
\label{metrics_more_epochs}
\resizebox{\textwidth}{!}{
\begin{tabular}{lcccccc}
\toprule
                  & \multicolumn{2}{c}{5 epochs}  & \multicolumn{2}{c}{10 epochs}   & \multicolumn{2}{c}{50 epochs} \\

                  & Train. set (75\%) & Val. set (25\%) & Train. set (75\%) & Val. set (25\%) & Train. set (75\%) & Val. set (25\%) \\
\midrule
Log-Loss          &    0.2931      &      0.3242         &   0.2364       &  0.3930  &          0.0663    &   1.2485         \\
NIR               &     0.8258          &        0.8265       &  0.8258       &   0.8265   &        0.8258 &       0.8265          \\
Accuracy          &    0.8294           &         0.8208      &    0.8313    &   0.8016    &      0.8597    &      0.7829         \\
TPR (Recall)      &   0.5718         &            0.5426   &     0.6626     &   0.5679    &        0.8190    &     0.5777        \\
TNR (Specificity) &   0.8838          &         0.8792        &   0.8668     &  0.8507     &      0.8682   &      0.8259        \\
Precision         &    0.5092           &      0.4854         &     0.5121         &  0.4441      & 0.5673     &  0.4107         \\
F\textsubscript{1} Score  & 0.5387     &        0.5124       &       0.5777             &  0.4984    &  0.6703     & 0.4801      \\
\bottomrule
\end{tabular}}
\end{table}

\subsection{Results for challenge submission}

For the test set, we are not able to compute the metrics defined in \autoref{metrics} as we do not have the labels. We can only indicate the best Log-Loss value obtained in the challenge: 6.622686 (Log-Loss value with transformation factor introduced by ShareChat RecSys Team to avoid cheating).

\subsection{Two outputs model: does it make a difference?}

We analyze if asking the model to predict one more output affects its ability to correctly predict the most interesting output ('is\_installed'). For that, we will compute the metrics described in \autoref{metrics}. The same metrics are given for both outputs in \autoref{metrics_two_outputs}. This experiment shows that we can achieve good results with the two outputs model and the values are similar to the ones in \autoref{metrics_is_installed}. For the 'is\_clicked' output, the metrics get values that are significantly worse than the one obtained for 'is\_installed'. This observation should be explained by the model not being sufficiently trained to correctly predict this output (we monitor the output 'is\_installed' to stop training). We will investigate that hypothesis in \autoref{predict_is_clicked}.

\begin{table}[]
\caption{Metrics for the training set and validation set (two outputs model)}
\label{metrics_two_outputs}
\begin{tabular}{lcccc}
\toprule
                  & \multicolumn{2}{c}{Output 'is\_clicked'}  & \multicolumn{2}{c}{Output 'is\_installed'}    \\

                  & Train. set (75\%) & Val. set (25\%) & Train. set (75\%) & Val. set (25\%) \\
\midrule
Log-Loss          &       0.3117        &      0.3192         &    0.3117          &     0.3191                   \\
NIR               &       0.7800        &      0.7807         &    0.8258          &     0.8265                  \\
Accuracy          &       0.6751        &      0.8217         &    0.8242          &     0.8213                  \\
TPR (Recall)      &       0.1909        &      0.5293         &    0.5440          &     0.5347                  \\
TNR (Specificity) &       0.8117        &      0.8831         &    0.8833          &     0.8815                  \\
Precision         &       0.2224        &      0.4874         &    0.4957          &     0.4865                  \\
F\textsubscript{1} Score        &       0.2054        &      0.5075         &    0.5187          &     0.5095                  \\
\bottomrule
\end{tabular}
\end{table}

\subsection{Predict 'is\_clicked'}
\label{predict_is_clicked}
Following the poor results we got for the 'is\_clicked' output using our two outputs model, we try to predict only 'is\_clicked' with the one output model defined in \autoref{one_output} in which we simply replace the output 'is\_installed' by the output 'is\_clicked'. \autoref{metrics_is_clicked} demonstrates that similar performance can be obtained for 'is\_clicked' (compared to the one output model predicting 'is\_installed'). The results in \autoref{metrics_two_outputs} were poor as the training was not completely done for the output 'is\_clicked' in the two outputs models. Indeed, training the model with only 'is\_clicked' as output required 4 epochs while all the other models required 3 epochs.

\subsection{A model to predict all the outputs}

Sharing all the layers between the inputs and the outputs (see \autoref{two_outputs}) seemed to be a good idea to share a maximum of information and also to avoid increasing the size of the model that we want to keep as small as possible. In fact, it is not a good idea as it doesn't allow the model to learn different representations of these two output variables.
The hypothesis made was that we could share knowledge as the output variables are similar but our experiments showed that we can't and that the outputs need to be treated independently. To use only one model, we need to duplicate all the layers that are used between the inputs and the outputs to allow the model to learn different weights values for each output. Moreover, we should stop training of each part of the model by monitoring the corresponding 'val\_loss'. This scenario is not so much better than using two models with one output, a solution that can also be used.

\begin{table}
\caption{Metrics for the training set and validation set (predict 'is\_clicked')}
\label{metrics_is_clicked}
\begin{tabular}{lcc}
\toprule
                  & Training set (75\%) & Validation set (25\%)   \\
\midrule
Log-Loss          				&    0.3373     &     0.3473              \\
NIR               				&    0.7800     &     0.7807              \\
Accuracy          				&    0.7873     &     0.7827              \\
TPR (Recall)      				&    0.6091     &     0.5963            \\
TNR (Specificity) 				&    0.8375     &     0.8350         \\
Precision         				&    0.5139     &     0.5038         \\
F\textsubscript{1} Score       &    0.5575     &     0.5462         \\
\bottomrule
\end{tabular}
\end{table}

\section{Conclusion}
\label{conclusion}

In challenges like the Recsys Challenge as in the daily life of researchers in machine learning and related fields, pursuing the best results based on a metric is often set as as goal. However, developping complex models to increase the 'prediction score' is not always necessary and the pros and cons should always be considered. With our model, we show that pretty good results can be achieved with a relatively simple model that can be easily used in many production environments. We also demonstrate that we can tune our model to achieve different results (monitor different metrics) and allow easy tuning following business needs.

\bibliographystyle{ACM-Reference-Format}
\bibliography{sample-base}


\begin{thebibliography}{11}


\ifx \showCODEN    \undefined \def \showCODEN     #1{\unskip}     \fi
\ifx \showDOI      \undefined \def \showDOI       #1{#1}\fi
\ifx \showISBNx    \undefined \def \showISBNx     #1{\unskip}     \fi
\ifx \showISBNxiii \undefined \def \showISBNxiii  #1{\unskip}     \fi
\ifx \showISSN     \undefined \def \showISSN      #1{\unskip}     \fi
\ifx \showLCCN     \undefined \def \showLCCN      #1{\unskip}     \fi
\ifx \shownote     \undefined \def \shownote      #1{#1}          \fi
\ifx \showarticletitle \undefined \def \showarticletitle #1{#1}   \fi
\ifx \showURL      \undefined \def \showURL       {\relax}        \fi
\providecommand\bibfield[2]{#2}
\providecommand\bibinfo[2]{#2}
\providecommand\natexlab[1]{#1}
\providecommand\showeprint[2][]{arXiv:#2}

\bibitem[Acock(2005)]%
        {mv1}
\bibfield{author}{\bibinfo{person}{Alan~C. Acock}.} \bibinfo{year}{2005}\natexlab{}.
\newblock \showarticletitle{Working With Missing Values}.
\newblock \bibinfo{journal}{\emph{Journal of Marriage and Family}} \bibinfo{volume}{67}, \bibinfo{number}{4} (\bibinfo{year}{2005}), \bibinfo{pages}{1012--1028}.
\newblock
\urldef\tempurl%
\url{https://doi.org/10.1111/j.1741-3737.2005.00191.x}
\showDOI{\tempurl}
\showeprint{https://onlinelibrary.wiley.com/doi/pdf/10.1111/j.1741-3737.2005.00191.x}


\bibitem[Brown and Kros(2003)]%
        {mv3}
\bibfield{author}{\bibinfo{person}{Marvin~L Brown} {and} \bibinfo{person}{John~F Kros}.} \bibinfo{year}{2003}\natexlab{}.
\newblock \showarticletitle{Data mining and the impact of missing data}.
\newblock \bibinfo{journal}{\emph{Industrial Management \& Data Systems}} \bibinfo{volume}{103}, \bibinfo{number}{8} (\bibinfo{year}{2003}), \bibinfo{pages}{611--621}.
\newblock


\bibitem[Canbek et~al\mbox{.}(2017)]%
        {metrics}
\bibfield{author}{\bibinfo{person}{Gürol Canbek}, \bibinfo{person}{Seref Sagiroglu}, \bibinfo{person}{Tugba~Taskaya Temizel}, {and} \bibinfo{person}{Nazife Baykal}.} \bibinfo{year}{2017}\natexlab{}.
\newblock \showarticletitle{Binary classification performance measures/metrics: A comprehensive visualized roadmap to gain new insights}. In \bibinfo{booktitle}{\emph{2017 International Conference on Computer Science and Engineering (UBMK)}}. \bibinfo{pages}{821--826}.
\newblock
\urldef\tempurl%
\url{https://doi.org/10.1109/UBMK.2017.8093539}
\showDOI{\tempurl}


\bibitem[Fire and Schler(2017)]%
        {fire_exploring_2017}
\bibfield{author}{\bibinfo{person}{Michael Fire} {and} \bibinfo{person}{Jonathan Schler}.} \bibinfo{year}{2017}\natexlab{}.
\newblock \showarticletitle{Exploring {Online} {Ad} {Images} {Using} a {Deep} {Convolutional} {Neural} {Network} {Approach}}. In \bibinfo{booktitle}{\emph{2017 {IEEE} {International} {Conference} on {Internet} of {Things} ({iThings}) and {IEEE} {Green} {Computing} and {Communications} ({GreenCom}) and {IEEE} {Cyber}, {Physical} and {Social} {Computing} ({CPSCom}) and {IEEE} {Smart} {Data} ({SmartData})}}. \bibinfo{pages}{1053--1060}.
\newblock
\urldef\tempurl%
\url{https://doi.org/10.1109/iThings-GreenCom-CPSCom-SmartData.2017.160}
\showDOI{\tempurl}


\bibitem[Pigott(2001)]%
        {mv4}
\bibfield{author}{\bibinfo{person}{Therese~D. Pigott}.} \bibinfo{year}{2001}\natexlab{}.
\newblock \showarticletitle{A Review of Methods for Missing Data}.
\newblock \bibinfo{journal}{\emph{Educational Research and Evaluation}} \bibinfo{volume}{7}, \bibinfo{number}{4} (\bibinfo{year}{2001}), \bibinfo{pages}{353--383}.
\newblock
\urldef\tempurl%
\url{https://doi.org/10.1076/edre.7.4.353.8937}
\showDOI{\tempurl}
\showeprint{https://doi.org/10.1076/edre.7.4.353.8937}


\bibitem[Qu et~al\mbox{.}(2016)]%
        {qu_product-based_2016}
\bibfield{author}{\bibinfo{person}{Yanru Qu}, \bibinfo{person}{Han Cai}, \bibinfo{person}{Kan Ren}, \bibinfo{person}{Weinan Zhang}, \bibinfo{person}{Yong Yu}, \bibinfo{person}{Ying Wen}, {and} \bibinfo{person}{Jun Wang}.} \bibinfo{year}{2016}\natexlab{}.
\newblock \showarticletitle{Product-{Based} {Neural} {Networks} for {User} {Response} {Prediction}}. In \bibinfo{booktitle}{\emph{2016 {IEEE} 16th {International} {Conference} on {Data} {Mining} ({ICDM})}}. \bibinfo{pages}{1149--1154}.
\newblock
\urldef\tempurl%
\url{https://doi.org/10.1109/ICDM.2016.0151}
\showDOI{\tempurl}
\newblock
\shownote{ISSN: 2374-8486}.


\bibitem[Raschka(2014)]%
        {raschka2014overview}
\bibfield{author}{\bibinfo{person}{Sebastian Raschka}.} \bibinfo{year}{2014}\natexlab{}.
\newblock \bibinfo{title}{An Overview of General Performance Metrics of Binary Classifier Systems}.
\newblock
\newblock
\showeprint[arxiv]{1410.5330}~[cs.LG]


\bibitem[Schafer(1999)]%
        {mv2}
\bibfield{author}{\bibinfo{person}{Joseph~L Schafer}.} \bibinfo{year}{1999}\natexlab{}.
\newblock \showarticletitle{Multiple imputation: a primer}.
\newblock \bibinfo{journal}{\emph{Statistical Methods in Medical Research}} \bibinfo{volume}{8}, \bibinfo{number}{1} (\bibinfo{year}{1999}), \bibinfo{pages}{3--15}.
\newblock
\urldef\tempurl%
\url{https://doi.org/10.1177/096228029900800102}
\showDOI{\tempurl}
\showeprint{https://doi.org/10.1177/096228029900800102}
\newblock
\shownote{PMID: 10347857}.


\bibitem[Ying(2019)]%
        {Ying_2019}
\bibfield{author}{\bibinfo{person}{Xue Ying}.} \bibinfo{year}{2019}\natexlab{}.
\newblock \showarticletitle{An Overview of Overfitting and its Solutions}.
\newblock \bibinfo{journal}{\emph{Journal of Physics: Conference Series}} \bibinfo{volume}{1168}, \bibinfo{number}{2} (\bibinfo{date}{feb} \bibinfo{year}{2019}), \bibinfo{pages}{022022}.
\newblock
\urldef\tempurl%
\url{https://doi.org/10.1088/1742-6596/1168/2/022022}
\showDOI{\tempurl}


\bibitem[Zhai et~al\mbox{.}(2016)]%
        {zhai_deepintent_2016}
\bibfield{author}{\bibinfo{person}{Shuangfei Zhai}, \bibinfo{person}{Keng-hao Chang}, \bibinfo{person}{Ruofei Zhang}, {and} \bibinfo{person}{Zhongfei~Mark Zhang}.} \bibinfo{year}{2016}\natexlab{}.
\newblock \showarticletitle{{DeepIntent}: {Learning} {Attentions} for {Online} {Advertising} with {Recurrent} {Neural} {Networks}}. In \bibinfo{booktitle}{\emph{Proceedings of the 22nd {ACM} {SIGKDD} {International} {Conference} on {Knowledge} {Discovery} and {Data} {Mining}}} \emph{(\bibinfo{series}{{KDD} '16})}. \bibinfo{publisher}{Association for Computing Machinery}, \bibinfo{address}{New York, NY, USA}, \bibinfo{pages}{1295--1304}.
\newblock
\showISBNx{978-1-4503-4232-2}
\urldef\tempurl%
\url{https://doi.org/10.1145/2939672.2939759}
\showDOI{\tempurl}


\bibitem[Zhang et~al\mbox{.}(2016)]%
        {zhang_deep_2016}
\bibfield{author}{\bibinfo{person}{Weinan Zhang}, \bibinfo{person}{Tianming Du}, {and} \bibinfo{person}{Jun Wang}.} \bibinfo{year}{2016}\natexlab{}.
\newblock \showarticletitle{Deep {Learning} over {Multi}-field {Categorical} {Data}}. In \bibinfo{booktitle}{\emph{Advances in {Information} {Retrieval}}} \emph{(\bibinfo{series}{Lecture {Notes} in {Computer} {Science}})}, \bibfield{editor}{\bibinfo{person}{Nicola Ferro}, \bibinfo{person}{Fabio Crestani}, \bibinfo{person}{Marie-Francine Moens}, \bibinfo{person}{Josiane Mothe}, \bibinfo{person}{Fabrizio Silvestri}, \bibinfo{person}{Giorgio~Maria Di~Nunzio}, \bibinfo{person}{Claudia Hauff}, {and} \bibinfo{person}{Gianmaria Silvello}} (Eds.). \bibinfo{publisher}{Springer International Publishing}, \bibinfo{address}{Cham}, \bibinfo{pages}{45--57}.
\newblock
\showISBNx{978-3-319-30671-1}
\urldef\tempurl%
\url{https://doi.org/10.1007/978-3-319-30671-1_4}
\showDOI{\tempurl}


\end{thebibliography}

\end{document}